\documentclass[%
 reprint,
 amsmath,amssymb,
 aps,
]{revtex4-1}

\usepackage{graphicx}
\usepackage{dcolumn}
\usepackage{bm}
\usepackage{longtable}
\usepackage{enumerate}
\usepackage{subfigure}
\usepackage{multirow}
\usepackage{cancel}
\usepackage{tabularx}
\usepackage{threeparttable}
\usepackage{dcolumn}
\usepackage{amssymb}
\usepackage{afterpage}
\usepackage{nicefrac}

\usepackage[version=3]{mhchem}
\usepackage{threeparttable}

\newcolumntype{d}[1]{D{.}{.}{#1}}

\newcommand{\Lower}[1]{\smash{\lower 1.5ex \hbox{#1}}}

\usepackage{ulem}

\newcommand{\mc}{\multicolumn{1}{c}}
\newcommand{\ml}{\multicolumn{1}{c|}}

\begin{document}

\title{Targeting excited states in all-trans polyenes with electron-pair states}
\author{Katharina Boguslawski}
\email{k.boguslawski@fizyka.umk.pl}
\affiliation{%
 Institute of Physics, Faculty of Physics, Astronomy and Informatics, Nicolaus Copernicus University in Torun, Grudzi{a}dzka 5, 87-100 Toru{n}, Poland \\
}%
\affiliation{%
 Faculty of Chemistry, Nicolaus Copernicus University in Torun, Gagarina 7, 87-100 Torun, Poland \\
}


\date{\today}

\begin{abstract}
 
Wavefunctions restricted to electron pair states are promising models for strongly-correlated systems.
Specifically, the pair Coupled Cluster Doubles (pCCD) ansatz allows us to accurately describe bond dissociation processes and heavy-element containing compounds with multiple quasi-degenerate single-particle states.
Here, we extend the pCCD method to model excited states using the equation of motion (EOM) formalism.
As the cluster operator of pCCD is restricted to electron-pair excitations, EOM-pCCD allows us to target excited electron-pair states only.
To model singly excited states within EOM-pCCD, we modify the configuration interaction ansatz of EOM-pCCD to contain also single excitations.
Our proposed model represents a simple and cost-effective alternative to conventional EOM-CC methods to study singly excited electronic states.
The performance of the excited state models is assessed against the lowest-lying excited states of the uranyl cation and the two lowest-lying excited states of all-trans polyenes.
Our numerical results suggest that EOM-pCCD including single excitations is a good starting point to target singly excited states.

\end{abstract}

\maketitle

\section{Introduction}

Quantum chemical modeling of electronic structures of excited states is usually more challenging than those of the ground state~\cite{bartlett_2007}. 
Specifically, we have to be able to describe various types of excited states simultaneously, each governed by different amounts of dynamic and nondynamic/static electron correlation effects~\cite{lee_1989,Bartlett_1994,entanglement_letter}.
Furthermore, excited states usually require larger one-electron basis functions that allow us to describe all correlation effects on an equal footing.  
The presence of minima, transition states, and surface crossings poses additional challenges when modeling their potential energy surfaces~\cite{SerranoAndrés200599}. 
Therefore, an accurate and reliable description of electronically excited states still remains an open problem in quantum chemistry. 

In the past two decades, many promising quantum chemistry approaches suitable for excited states have emerged (see, for instance, recent reviews~\cite{piecuch2002review,SerranoAndrés200599,excitations-review,spin-flip-EOM,Roos_rev_2008,nishikawa2012preface}).
In wavefunction-based methods, the most popular approaches are the multi-reference configuration interaction, complete active space self-consistent field theory~\cite{Roos_casscf,Siegbahn_casscf}, including dynamic energy corrections~\cite{MS-CASPT2,caspt21,nevpt2,Pulay_CASPT2_2011,RASPT2}, and coupled cluster ans\"atze. 
Specifically, in coupled cluster theory, considerable development was done in the Fock-Space coupled cluster (FSCC) formulation~\cite{meissner-musial-chapter,monika_mrcc} and the equation of motion formalism~\cite{Rowe-EOM,Dunning-EOM,jankowski-eom,Shavitt_book,Helgaker_book}. 
In the standard formulation of the FSCC approach, one starts from a closed-shell electronic configuration described by a standard coupled cluster approach and calculates the electronically excited states using the information from ionization potentials and electron affinities. 
Alternatively, electronic spectra can be obtained from the high-spin reference wavefunction. 
Despite its sound theoretical basis, the FSCC approach suffers form convergence difficulties. 
To remedy this problem, intermediate Hamiltonian techniques have been developed~\cite{IHFSCC_11,IHFSCC_2,IHFSCC_3,FSCC_Leszek_2008,FSCC_02_sector}, but convergence difficulties still remain for large and complex molecular systems~\cite{pawel3}. 
In the equation of motion coupled cluster (EOM-CC) method, one obtains electronic spectra by introducing a linear excitation operator on top of the CC ground state wavefunction.
This theory has been successfully applied to various problems in chemistry and physics~\cite{EOMCCSd,EOMCC_2001,Kowalski_fused_porphyrine,Lopata_2011,pawel_saldien,SOC-EOM}. 
However, the standard EOM-CC method does not produce accurate excitation energies for systems with significant multi-reference character. 
This deficiency led to the development of its modified variants, such as, the spin-flip EOM-CC~\cite{spin-flip-EOM-theory,spin-flip-EOM}, completely renormalized EOM-CC~\cite{CR-EOMCCSD,unrestricted_CR-EOMCCSD,Kowalski_porhyrin} and active space EOM-CC methods~\cite{kowalski2000active,kowalski2005active,piecuch2010active,kowalski2010active,2-2-EOM-theory,CR-2-2-EOM-theory}. 
Large-scale modeling of excited states using the EOM-CC formalism has been available with the NWChem software package~\cite{kowalski2011scalable,van2011nwchem}. 
Simplified EOM-CC models represent alternative approaches to describe excited states in large molecular system~\cite{GWALTNEY1996189}. 
Another example is the recently formulated equation of motion linear coupled cluster ansatz~\cite{EOM-LCC}. 

In this work we derive and implement another simplified version of the EOM method to describe electronically excited states in large molecular systems. 
Specifically, we use the pair coupled cluster doubles (pCCD) ansatz~\cite{Limacher_2013,OO-AP1roG} to model the initial ground-state electronic structure.
Recent applications of this model to ground-state properties, including bond-breaking processes, have been very encouraging ~\cite{Limacher_2013,OO-AP1roG,pawel_jpca_2014,Piotrus_Mol-Phys,PS2-AP1roG,AP1roG-JCTC,pawel_PCCP2015,Limacher_2015,Piotrus_PT2,Kasia-lcc,garza2015actinide,
henderson2015pair,gomez2016singlet,Boguslawski2016}. 
The pCCD wavefunction is a simplified CC-type wavefunction, where the cluster operator is restricted to electron pair excitations~\cite{p-CCD,Tamar-pCC},
\begin{equation}\label{eq:ap1rog}
|{\rm pCCD}\rangle = \exp \left (  \sum_{i=1}^P \sum_{a=P+1}^K c_i^a a_{a \uparrow}^{\dagger}  a_{a\downarrow}^{\dagger} a_{i\downarrow}  a_{i\uparrow}\right )|0 \rangle,
\end{equation}
where $|0 \rangle$ is some reference determinant. Indices $i$ and $a$ correspond to occupied and virtual orbitals with respect to $|0 \rangle$, $P$ and $K$ denote the number of electron pairs and orbitals, respectively.
The pCCD ansatz is also known as the antisymmetric product of 1-reference orbital geminal (AP1roG) ansatz, where the pair-coupled-cluster amplitudes correspond to the geminal coefficients.
This wavefunction ansatz is, by construction, size-extensive and has mean-field scaling if the geminal coefficients/amplitudes are optimized using the projected Schr\"odinger equation approach. Note that $|0 \rangle$ is usually optimized as well and hence differs from the Hartree--Fock determinant.

This work is organized as follows.
In section~\ref{sec:theory}, we briefly summarize the equation of motion formalism with a pCCD reference function as well as a simple and cost-effective extension to account for single excitations.
Numerical examples are presented in section~\ref{sec:results}, where we investigated the lowest lying excited states of the \ce{UO2^{2+}} molecule as a test case and the two lowest-lying excited states of all-trans polyenes.
Finally, we conclude in section~\ref{sec:conclusions}.

\section{Targeting excited states in pCCD}\label{sec:theory}
In order to model excited states in pCCD, we will use the equation of motion (EOM) formalism~\cite{bartlett-eom,jankowski-eom,bartlett_2007}.
In EOM-CC, excited states are parametrized by a linear CI-type ansatz~\cite{Helgaker_book},
\begin{equation}
        \hat{R} = \sum_\mu c_\mu \hat{\tau}_\mu,
\end{equation}
where the summation is over all excitation operators present in the cluster operator $\hat{T}$ as well as the identity operator $\hat{\tau}_0$.
The operator $\hat{R}$ is then used to generate the target state from the initial CC state,
\begin{equation}
        \vert \Psi \rangle = \hat{R} \exp(\hat{T}) | 0 \rangle = \sum_\mu {c_\mu \hat{\tau}_\mu} \exp(\hat{T}) | 0 \rangle,
\end{equation}
with $\vert 0 \rangle$ being the CC reference determinant.

To arrive at the EOM-CC working equations, it is convenient to use the normal-product form of the Hamiltonian, $\hat{H}_N = \hat{H} - \langle 0 \vert \hat{H} \vert 0 \rangle $, written as a sum of the Fock operator $\hat{F}_N$ and the electron repulsion term $\hat{W}_N$ (in physicists' notation), 
\begin{equation}
        \hat{H}_N = \sum_{pq} f_{pq} \{a_p^\dagger a_q\} + \frac{1}{2} \sum_{pqrs} \langle pq \vert rs \rangle \{ a_p^\dagger a_q^\dagger a_s a_r \}.
\end{equation}
Furthermore, we will disregard any excitation properties, like dipole moments, and focus on excitation energies instead.
In that case, we have to solve for the $\hat{R}$ amplitudes only.
Our target-state Schr\"odinger equation then reads
\begin{equation}\label{eq:tse}
        \hat{H}_N \hat{R} \exp{(\hat{T})} \vert 0 \rangle = \Delta E \hat{R} \exp{(\hat{T})} \vert 0 \rangle,
\end{equation}
where $\Delta E$ is the energy difference with respect to the Fermi vacuum expectation value $\vert 0 \rangle$.
Introducing the similarity transformed Hamiltonian in normal-product form $\hat{\mathcal{H}}_N = \exp{(-\hat{T})} \hat{H}_N \exp{(\hat{T})}$ and subtracting the equation for the CC ground state, we obtain the EOM-CC equations for the $\hat{R}$ amplitudes,
\begin{equation}\label{eq:eomcc}
        [\hat{\mathcal{H}}_N,\hat{R}] \vert 0 \rangle  = \omega \hat{R} \vert 0 \rangle,
\end{equation}
where $\omega$ are the excitation energies with respect to the CC ground state, $ \exp(\hat{T}) | 0 \rangle$.
The excitation energies are thus the eigenvalues of a non-Hermitian matrix,
\begin{equation}
        \begin{bmatrix}
          0 & \langle 0 \vert \hat{\mathcal{H}}_N \vert \mu \rangle \\
          0 & \langle \nu \vert [\hat{\mathcal{H}}_N,\tau_\mu ] \vert 0 \rangle 
        \end{bmatrix},
\end{equation}
where the first row is associated with the CC reference state and the subsequent rows correspond to the excited configurations $\nu > 0$.
The EOM-CC working equations may be solved using, for instance, non-Hermitian extensions of the Davidson algorithm to determine the lowest-lying excited electronic states.

\subsection{Electron-pair excitations}
In this work, we are considering an pCCD reference function $\vert \mathrm{pCCD} \rangle$ as a special CC state confined to electron-pair states.
As indicated in eq.~\eqref{eq:ap1rog}, the pCCD cluster operator contains only electron-pair excitations, $\hat{T}=\hat{T}_p= \sum_{ia} t_i^a a^\dagger_a a^\dagger_{\bar{a}} a_{\bar{i}} a_i$.
In the corresponding EOM model, the $\hat{R}$ operator is thus restricted to the identity operator $\hat{\tau}_0$ as well as all pair excitations present in the cluster operator $\hat{T}_p$,
\begin{equation}\label{eq:rp}
        \hat{R}_p =  c_0 \hat{\tau}_0 + \sum_{ia} c_{i\bar{i}}^{a\bar{a}} \hat{\tau}_{a\bar{a}i\bar{i}},
\end{equation}
where $\hat{\tau}_{a\bar{a}i\bar{i}}=a^\dagger_a a^\dagger_{\bar{a}} a_{\bar{i}} a_i$ creates an electron pair in the virtual orbital $a$.
To obtain the target-state Schr\"odinger equation of EOM-pCCD, we have to substitute the general cluster operator $\hat{T}$ by the pair-excitation operator $\hat{T}_p$ in eq.~\eqref{eq:tse}, 
\begin{equation}
        \hat{H}_N \hat{R}_p \exp{(\hat{T}_p)} \vert 0 \rangle = \Delta E \hat{R}_p \exp{(\hat{T}_p)} \vert 0 \rangle.
\end{equation}
The $\hat{R}_p$ amplitudes are determined from the EOM-pCCD equations (restricting $\hat{R}$ to $\hat{R}_p$ and $\hat{T}$ to $\hat{T}_p$ in eq.~\eqref{eq:eomcc}),
\begin{equation}
        [\hat{\mathcal{H}}^{(p)}_N,\hat{R}_p] \vert 0 \rangle  = \omega_p \hat{R}_p \vert 0 \rangle,
\end{equation}
where $\mathcal{\hat{H}}^{(p)}_N $ indicates the similarity transformed Hamiltonian of pCCD, $ \mathcal{\hat{H}}^{(p)}_N = \exp{(-\hat{T}_P)} \hat{H}_N \exp{(\hat{T}_P)}$, and $\omega_p$ are the electron-pair excitation energies.
Thus, EOM-pCCD allows us to model electron-pair excited states only.
In order to target singly excited or general doubly excited states, we have to extend the pCCD cluster operator to include excitations beyond electron pairs.
This can be done either by changing to a frozen-pair CCSD~\cite{frozen-pCCD} formalism or to a linearized CCSD correction with an pCCD reference function~\cite{Kasia-lcc}.
In this work however, we will consider a different, cost-effective approach to account for single excitations in the pCCD model that does not scale as $\mathcal{O}(n^6)$ as conventional EOM-CCSD methods.
Note that EOM-pCCD scales as $\mathcal{O}(o^2v^2)$, where $o$ is the number of occupied orbitals (equivalent to the number of electron pairs) and $v$ is the number of virtual orbitals.

\subsection{Accounting for single excitations}
As proposed by Forseman \textit{et al.}~\cite{foresman-CIS}, configuration interaction with only single substitutions (CIS) represents an accurate model to investigate (singly) excited electronic states, even for large systems.
In the CIS method, the reference is a single Slater determinant obtained from an SCF procedure, while the CIS wavefunction is expanded as
\begin{equation}
    \vert \mathrm{CIS} \rangle = c_0 \vert 0 \rangle + \sum_{ia} c_{i}^{a} \vert _i^a \rangle,
\end{equation}
with $\vert _i^a \rangle$ being a singly excited determinant where the (occupied) orbital $i$ of $\vert 0 \rangle$ has been substituted by the (virtual) orbital $a$.
Similar to CIS, we will include single excitations in EOM-pCCD by extending the $\hat{R}_p$ operator of eq.~\eqref{eq:rp}.
In addition to the identity operator $\hat{\tau}_0$ and all pair excitations $\hat{T}_p$, $\hat{R}_p$ also contains a summation over all single excitations,
\begin{equation}
    \hat{R}_{ps} =  c_0 \hat{\tau}_0 + \sum_{ia} c_{i}^{a} \hat{\tau}_{ai} + \sum_{ia} c_{i\bar{i}}^{a\bar{a}} \hat{\tau}_{a\bar{a}i\bar{i}},
\end{equation}
where $\hat{\tau}_{ai}$ is a singlet excitation operator $ \hat{\tau}_{ai} = a^\dagger_a  a_i  +  a^\dagger_{\bar{a}} a_{\bar{i}}$ that creates a singly excited electronic state with respect to $\vert 0 \rangle$, $\vert _i^a \rangle = \hat{\tau}_{ia} \vert 0 \rangle$.
The $ \hat{R}_{ps}$ amplitudes are determined from solving 
\begin{equation}
        [\hat{\mathcal{H}}^{(p)}_N,\hat{R}_{ps}] \vert 0 \rangle  = \omega_{ps} \hat{R}_{ps} \vert 0 \rangle,
\end{equation}
where we still have the similarity transformed Hamiltonian of pCCD, $\mathcal{\hat{H}}^{(p)}_N $, while $\omega_{ps}$ are the excitation energies of both singly excited and pair excited states.
We will label this simplified model as EOM-pCCD+S to indicate that single excitations are included \textit{a posteriori} in the $\hat{R}_p$ operator.
In contrast to CIS that uses a single Slater determinant as reference, EOM-pCCD+S employs the pCCD wavefunction as reference state.
Furthermore, electron correlation effects are included through the $\hat{T}_p$ operator in the similarity transformed Hamiltonian.

Similar to EOM-pCCD, the excitation energies are obtained by diagonalizing a non-Hermitian matrix of the form
\begin{equation}
        \begin{bmatrix}
          0 & \langle 0 \vert \hat{\mathcal{H}}_N^{(p)} \vert _j^b \rangle & \langle 0 \vert \hat{\mathcal{H}}_N^{(p)} \vert _{j\bar{j}}^{b\bar{b}} \rangle \\
          \langle _i^a \vert \hat{\mathcal{H}}_N^{(p)} \vert 0 \rangle &  \langle _i^a \vert \hat{\mathcal{H}}_N^{(p)} \vert _{j}^{b} \rangle & \langle _i^a \vert \hat{\mathcal{H}}_N^{(p)} \vert _{j\bar{j}}^{b\bar{b}} \rangle \\
          0 & \langle _{i\bar{i}}^{a\bar{a}} \vert \hat{\mathcal{H}}_N^{(p)} \vert _{j}^{b} \rangle & \langle _{i\bar{i}}^{a\bar{a}} \vert \hat{\mathcal{H}}_N^{(p)} \vert _{j\bar{j}}^{b\bar{b}} \rangle \\
        \end{bmatrix}.
\end{equation}
Note that in contrast to conventional EOM-CC methods, the first column does not equal zero because single excitations are not included in the cluster operator of pCCD.
Thus, terms like $\langle _i^a \vert \hat{\mathcal{H}}_N^{(p)} \vert 0 \rangle$ do not vanish as they are not incorporated in the ground-state CC amplitude equations.
Although we can account for single excitations in a rather straightforward way, we loose size-intensivity in the EOM model.
For the molecular systems investigated in this work, the error introduced by extending only the $\hat{R}_p$ operator is approximately three orders of magnitude smaller than the actual excitation energies, while the computational cost increases insignificantly compared to EOM-pCCD (the Hamiltonian still contains terms that scale as $\mathcal{O}(o^2v^2)$, but with a larger pre-factor).
Thus, EOM-pCCD+S represents a cost-effective starting point to study singly excited electronic states in the pCCD model.

\begin{figure*}[htbp]
\centering
\includegraphics[width=0.99\linewidth]{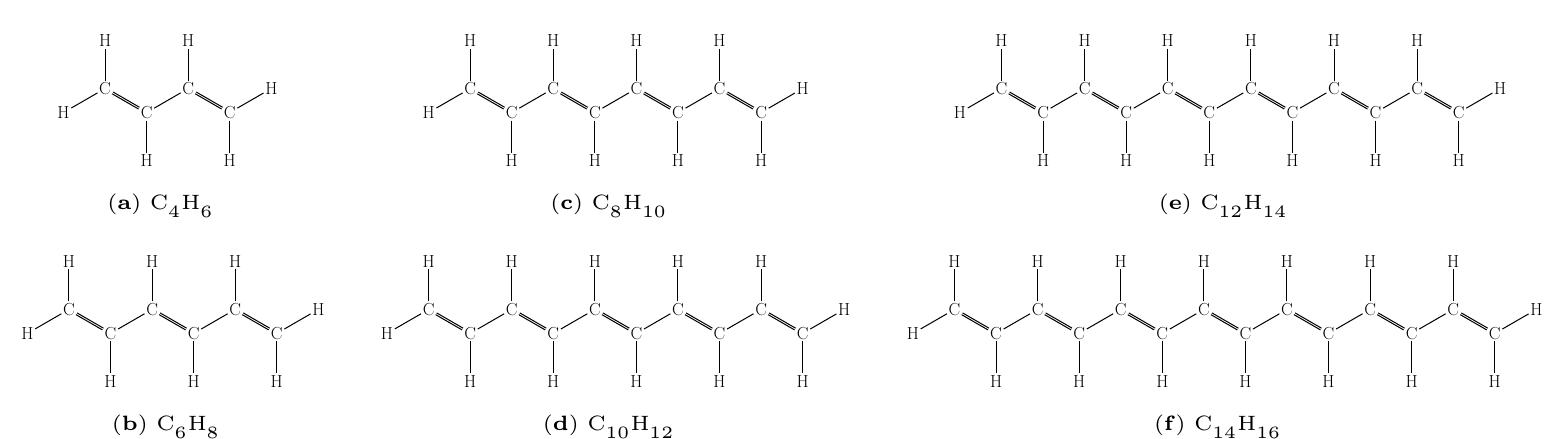}
\caption{Lewis structures of all investigated all-trans polyenes.
}
\label{fig:structures}
\end{figure*}

\section{Computational details}\label{sec:compdetails}

\subsection{Molecular structures}

In this work, we investigate all-trans polyenes containing 2 to 7 $\pi$-bonds (the corresponding Lewis structures are shown in Figure~\ref{fig:structures}).
Specifically, we study five different geometries, three for the lowest lying vertically excited states and two for lowest lying adiabatically excited states.
The first set of structures was optimized for each ground state using the ADF software package~\cite{adf2014,adf1,adf2} with a DZ basis set~\cite{adf_b} and the B3LYP exchange--correlation functional~\cite{B3LYP}.
The remaining sets of geometries were taken from ref.~\cite{DMRG-geom-opt} and were optimized using the density matrix renormalization group (DMRG).
Those include molecular geometries for the vertically and adiabatically excited states of \ce{C10H12}, \ce{C12H14}, and \ce{C14H16}.
Furthermore, for both the vertically and adiabatically excited states, we study geometries optimized using two different active spaces as discussed in ref.~\cite{DMRG-geom-opt}: an active space containing only $\pi$-electrons (abbreviates as $v_\pi$ and $a_\pi$) and an active space comprising both $\sigma$- and $\pi$-electrons (denoted by $v$ and $a$, respectively).

For the linear \ce{UO_2^{2+}} molecule, we calculated the four lowest-lying vertically and adiabatically excited states.
The position of the minimum of each potential energy surface was obtained by simultaneously stretching the U--O bond between 1.610 and 1.870 \AA{}.

\subsection{Calculation of ground and excited states}
The pCCD, EOM-pCCD, and EOM-pCCD+S calculations were performed with our in-house quantum-chemistry code employing restricted canonical Hartree--Fock orbitals as molecular orbital basis.
The lowest-lying excitation energies were determined employing a modified Davidson algorithm.
We should emphasize that the molecular orbital basis was not optimized within the pCCD method.
Orbital optimization in pCCD typically results in localized, symmetry-broken orbitals that preclude us from identifying the symmetry of excited states.
Furthermore, orbital optimization is only performed for the pCCD ground state wavefunction and results in molecular orbitals that are biased toward the ground state.
This bias can be reduced by introducing a state-average protocol yielding a set of compromise orbitals (as usually done in multi-configuration SCF methods).
Since we need to identify the symmetry of the (lowest-lying) excited states, we chose to not optimize the orbital basis and to use orbitals that transform according to the irreducible representations of the molecular point group (numerical examples for excitation energies within pCCD obtained in the optimized orbital basis can be found in the Supporting Information).
Moreover, recent numerical studies confirm that pCCD using restricted Hartree--Fock orbitals captures, at least qualitatively, the most important correlation effects around the equilibrium geometry~\cite{Boguslawski2016}.

For the investigated all-trans polyenes, we used the 6-31G basis set as well as the cc-pVDZ basis set of Dunning~\cite{dunning_b} for direct comparison with either CIS(D) (6-31G) or MRMP/DMRG (cc-pVDZ) reference data, respectively.

For \ce{UO_2^{2+}}, scalar relativistic effects were incorporated through relativistic effective core potentials (RECP). In all calculations, we have used a small core (SC) RECP (60 electrons in the core) with the following contraction scheme (12s11p10d8f) $\rightarrow$ [8s7p6d4f]~\cite{U_RLC}. For the lighter elements (O), the cc-pVDZ basis set of Dunning~\cite{dunning_b} was employed, (10s5p1d) $\rightarrow$ [4s3p1d].
The CIS and EOM-CCSD calculations have been performed with the \textsc{Molpro} software package~\cite{molpro2012}.
No frozen core was used in all calculations.

\section{The UO$_2^{2+}$ molecule as a test case}\label{sec:results-test}

Before scrutinizing the performance of EOM-pCCD and EOM-pCCD+S in modeling the lowest-lying excited states in all-trans polyenes, we will assess the accuracy of EOM-pCCD+S against molecular systems whose lowest-lying excited states are purely singly-excited states.
For that purpose, we chose the uranyl cation (\ce{UO_2^{2+}}) as it contains a large number of strongly-correlated electrons distributed among the 5$f$-, 6$d$-, and 7$s$-orbitals and its electronic spectrum is well understood~\cite{uranyl_pitzer_99,pierloot05,denning07,real07,pawel1,pawel2}.

The vertical and adiabatic excitation energies for the four lowest-lying excited states in EOM-pCCD+S and CIS are summarized in Table~\ref{tab:uo2}.
Note that for the \ce{UO_2^{2+}} molecule the two lowest-lying excited states can be accurately described within the EOM-CCSD model, which yields excitations energies that are similar to completely renormalized EOM-CCSD(T) reference values~\cite{pawel_saldien}.
For the higher-lying excited states, however, dynamic correlation effects become important and a triples correction has to be included to accurately model those states.
The EOM-CCSD results in Table~\ref{tab:uo2} can thus be considered as upper bounds of the excitation energies for the $\pi_u \rightarrow \delta_u$ state.

In general, EOM-pCCD+S overestimates vertical excitations energies of the two lowest-lying excited states ($\sigma_u\rightarrow \phi_u$ and $\sigma_u\rightarrow \delta_u$) by approximately 0.5 eV, while the corresponding adiabatic excitation energies are overestimated by about 0.7 eV.
Note that for the uranyl cation, CIS outperforms EOM-pCCD+S and deviates from EOM-CCSD reference data by approximately 0.3 (vertical excitations) to 0.6 eV (adiabatic excitations).
As expected, neither EOM-pCCD+S nor CIS are able to accurately predict the excitation energies for the $\pi_u \rightarrow \delta_u$ states as these excited states are dominated by dynamic correlation effects which are not included in CIS and only marginally accounted for in pCCD (see also~\cite{Boguslawski2016}).
We should emphasize that EOM-pCCD+S provides equilibrium U--O bond lengths of excited states that deviate less from the EOM-CCSD reference values ($\Delta r_e \approx 0.06$ \AA{} compared to $\Delta r_e \approx 0.08$ \AA{} in CIS).

To conclude, the EOM-pCCD+S model predicts the correct order of the lowest-lying excited states in the uranyl cation and provides excitation energies of decent accuracy with errors of about 0.5 eV with respect to EOM-CCSD reference values.
However, errors in excitation energies are slightly worse than in the simple CIS model.
\begin{table*} 
\begin{center}
\caption{Vertical and adiabatic excitation energies and equilibrium U--O bond lengths of the four lowest-lying excited states in the \ce{UO2^{2+}} molecule calculated for EOM-pCCD+S and different quantum chemistry methods.
}\label{tab:uo2}
\begin{tabular}{l| d{3}d{3}d{3}d{3} | d{3}d{3}d{3}d{3} | d{4}d{4}d{4}d{4} }
\hline
\hline
\multirow{3}{*}{method} & \multicolumn{4}{c|}{vertical} & \multicolumn{8}{c}{adiabatic} \\ \cline{2-13}
 & \multicolumn{4}{c|}{$\omega$ [eV]} & \multicolumn{4}{c|}{$\omega$ [eV]} & \multicolumn{4}{c}{$r_e$ [\AA{}]} \\ \cline{2-13}
 & \multicolumn{1}{c}{$\sigma_u\rightarrow \phi_u$} & \multicolumn{1}{c}{$\sigma_u\rightarrow \delta_u$} & \multicolumn{1}{c}{$\pi_u \rightarrow \delta_u$} & \multicolumn{1}{c|}{$\pi_u \rightarrow \delta_u$} & 
   \multicolumn{1}{c}{$\sigma_u\rightarrow \phi_u$} & \multicolumn{1}{c}{$\sigma_u\rightarrow \delta_u$} & \multicolumn{1}{c}{$\pi_u \rightarrow \delta_u$} & \multicolumn{1}{c|}{$\pi_u \rightarrow \delta_u$} & 
   \multicolumn{1}{c}{$\sigma_u\rightarrow \phi_u$} & \multicolumn{1}{c}{$\sigma_u\rightarrow \delta_u$} & \multicolumn{1}{c}{$\pi_u \rightarrow \delta_u$} & \multicolumn{1}{c}{$\pi_u \rightarrow \delta_u$} \\ \hline
EOM-pCCD+S & 4.45  & 4.83  & 7.79  & 7.93  & 4.43  & 4.82  & 7.63  & 7.80  & 1.711 & 1.709 & 1.746 & 1.744\\
CIS        & 4.33  & 4.79  & 7.84  & 8.00  & 4.18  & 4.66  & 7.47  & 7.62  & 1.694 & 1.690 & 1.719 & 1.720\\
EOM-CCSD   & 4.02  & 4.36  & 5.28  & 5.36  & 3.70  & 4.08  & 4.63  & 4.72  & 1.772 & 1.768 & 1.805 & 1.805\\
\hline
\hline
\end{tabular}
\end{center}
\end{table*} 

\begin{table*} 
\begin{center}
\caption{Vertical excitation energies of the two lowest-lying excited states in all-trans polyenes \ce{C4H6} to \ce{C14H16} calculated for EOM-pCCD, EOM-pCCD+S, and different quantum chemistry methods.
Note that the 6-31G basis set was used in CIS(D), while the cc-pVDZ basis set was utilized in MRMP. CASSCF was performed in a double-zeta basis set (see corresponding references).
The excitation energies of EOM-pCCD and EOM-pCCD+S are determined for the cc-pVDZ basis set, while the corresponding results for the 6-31G basis set are given in parenthesis.
}\label{tab:dft}
\begin{tabular}{c| d{8}d{10}d{4}d{4}d{4}|d{10}d{4}d{4}d{4}}
\hline
\hline
& \multicolumn{5}{c|}{$2^1A_g^-$} & \multicolumn{4}{c}{$1^1B_u^+$} \\ \cline{2-10}
\ce{C=C} & \multicolumn{1}{c}{EOM-pCCD} & \multicolumn{1}{c}{EOM-pCCD+S} & \multicolumn{1}{c}{CIS(D)\footnotemark[1]} & \multicolumn{1}{c}{MRMP\footnotemark[2]}  & \multicolumn{1}{c|}{CASSCF\footnotemark[3]} & \multicolumn{1}{c}{EOM-pCCD+S} & \multicolumn{1}{c}{CIS(D)\footnotemark[1]} & \multicolumn{1}{c}{MRMP\footnotemark[2]} & \multicolumn{1}{c}{CASSCF\footnotemark[3]} \\\hline
2 & 10.56\,(10.49)& 7.45\,(7.37) & 9.01 & 6.31  & 6.67  & 7.20\,(7.44) & 8.09  & 6.21  & 7.73  \\
3 & 9.11\,(9.04)  & 6.79\,(6.75) & 7.81 & 5.10  & 5.64  & 5.98\,(6.16) & 6.78  & 5.25  & 7.06  \\
4 & 8.11\,(8.02)  & 6.15\,(6.09) & 6.78 & 4.26  & 5.16  & 5.19\,(5.34) & 5.95  & 4.57  & 6.62  \\
5 & 7.42\,(7.32)  & 5.69\,(5.63) & 6.12 & 3.68  & 4.32  & 4.62\,(4.75) & 5.43  & 4.17  & 6.37  \\
6 & 6.93\,(6.82)  & 5.37\,(5.29) & 5.55 & 3.19  &\ml{--}& 4.20\,(4.31) & 5.00  & 3.87  &\mc{--}\\
7 & 6.58\,(6.46)  & 5.13\,(5.05) & 5.14 & 2.80  &\ml{--}& 3.87\,(3.97) & 4.70  & 3.60  &\mc{--}\\
\hline
\hline
\end{tabular}
\footnotetext[1]{Taken from Ref.~\cite{Starcke2006}}
\footnotetext[2]{Taken from Ref.~\cite{Kurashige2004}}
\footnotetext[3]{Taken from Ref.~\cite{Nakayama}}
\end{center}
\end{table*} 

\begin{table*} 
\begin{center}
\caption{Vertical and adiabatic excitation energies of the two lowest-lying excited states in all-trans polyenes \ce{C10H12} to \ce{C14H16} calculated with EOM-pCCD, EOM-pCCD+S, and DMRG for different DMRG-optimized geometries.
Note that different active spaces are used in DMRG calculations (see computational details and ref.~\cite{DMRG-geom-opt}), while all orbitals are active in EOM-pCCD and EOM-pCCD+S.
The DMRG reference data is taken from ref.~\cite{DMRG-geom-opt}.
Experimental data is taken from ref.~\cite{Kohler}.
}\label{tab:dmrg}
\begin{tabular}{c| d{7}d{3}|d{7}d{3}|d{7}d{3}|d{7}d{3}|d{3}}
\hline
\hline
& \multicolumn{9}{c}{$2^1A_g^-$} \\ \cline{2-10}
&  \multicolumn{2}{c|}{$v$} & \multicolumn{2}{c|}{$a$} &  \multicolumn{2}{c|}{$v_\pi$} &  \multicolumn{2}{c|}{$a_\pi$} \\ \cline{2-9}
\ce{C=C} & \multicolumn{1}{c}{EOM-pCCD+S} & \multicolumn{1}{c|}{DMRG}
         & \multicolumn{1}{c}{EOM-pCCD+S} & \multicolumn{1}{c|}{DMRG}
         & \multicolumn{1}{c}{EOM-pCCD+S} & \multicolumn{1}{c|}{DMRG}
         & \multicolumn{1}{c}{EOM-pCCD+S} & \multicolumn{1}{c|}{DMRG} & \mc{Exp.} \\\hline
5 & 6.28 & 5.43 & 5.18 & 4.01 & 6.07 & 4.51 & 4.85 & 3.36 & 3.03\\
6 & 5.99 & 4.76 & 4.61 & 3.41 & 5.84 & 4.15 & 4.60 & 2.99 & 2.69\\
7 & 5.76 & 4.64 & 4.44 & 3.22 & 5.63 & 3.91 & 4.45 & 2.73 & 2.44\\ \hline \hline
         & \multicolumn{9}{c}{$1^1B_u^+$} \\ \cline{2-10}
&  \multicolumn{2}{c|}{$v$} & \multicolumn{2}{c|}{$a$} &  \multicolumn{2}{c|}{$v_\pi$} &  \multicolumn{2}{c|}{$a_\pi$} \\ \cline{2-9}
         & \multicolumn{1}{c}{EOM-pCCD+S} & \multicolumn{1}{c|}{DMRG}
         & \multicolumn{1}{c}{EOM-pCCD+S} & \multicolumn{1}{c|}{DMRG}
         & \multicolumn{1}{c}{EOM-pCCD+S} & \multicolumn{1}{c|}{DMRG}
         & \multicolumn{1}{c}{EOM-pCCD+S} & \multicolumn{1}{c|}{DMRG} & \mc{Exp.}\\\cline{2-10}
5 & 4.91 & 5.35 & 5.18 & 4.98 & 4.79 & 5.77 & 4.62 & 5.49 & 3.57\\
6 & 4.50 & 4.98 & 4.30 & 4.60 & 4.41 & 5.41 & 4.26 & 5.13 & 3.31\\
7 & 4.25 & 4.66 & 4.04 & 4.29 & 4.14 & 5.16 & 3.99 & 4.87 & 3.12\\
\hline
\hline
\end{tabular}
\end{center}
\end{table*} 

\section{Excited states in all-trans polyenes}\label{sec:results}

All-trans polyenes are model systems for carotenoids and polyene chromophores that play an important role in photoprocesses.
The proper description of the two lowest-lying excited states poses a challenge to both experiment and quantum chemistry approaches~\cite{Hudson1972,Schulten1976,Cave1987,Tavan1987,Cave1988,Buma1991,Orlandi1991,Serrano1993a,Serrano1993b,Krawczyk2000,Hsu2001,Wiberg2002,Starcke2006,polyeneLimit}, especially because doubly excited configuration are required to accurately model ground and excited states of longer polyenes.
The excitation energies for the two-lowest vertical excited states of \ce{C4H6} to \ce{C14H16} using the DFT optimized structures and two different basis sets (cc-pVDZ and 6-31G) are summarized in Table~\ref{tab:dft}.
The excitation energies obtained in the 6-31G basis are given in parenthesis next to the corresponding results calculated for the cc-pVDZ basis set.
Note that the 6-31G basis set was used in CIS(D)~\cite{Starcke2006}, while a cc-pVDZ basis set was employed in MRMP calculations ~\cite{Kurashige2004}.
Since EOM-pCCD can only model pair excited states, only the excitation energies of the $2^1A_g^-$ state are shown in the Table.

For all investigated polyenes, EOM-pCCD overestimates the excitation energies of the first excited state $2^1A_g^-$ by approximately 4 eV compared to MRMP and 1.5 eV compared to CIS(D) data.
This error can be partially attributed to missing single excitations that have to be included in the model to describe the character of the $2^1A_g^-$ state accurately.
Including singles on top of EOM-pCCD \textit{a posteriori} reduces the error by approximately 2 to 1 eV, depending on the number of $\pi$ bonds.
While EOM-pCCD+S predicts excitation energies that are lower than the corresponding excitation energies of CIS(D) (differences amount up to 1.7 eV), it overestimates the excitation energies of the $2^1A_g^-$ state by about 2 eV compared to MRMP data.
Furthermore, the simple EOM-pCCD+S model does not predict the right order of states where the first dark state $2^1A_g^-$ should lie below the first bright state $1^1B_u^+$.
The excitation energies of the first bright state $1^1B_u^+$, however, deviate only by approximately 0.3 eV from MRMP data.
We should note that EOM-pCCD+S outperforms CIS(D) in predicting the excitation energies of the $1^1B_u^+$ state.
Despite its simplicity, EOM-pCCD+S is thus a good starting point to model singly excited states.

\begin{figure}[htbp]
\centering
\includegraphics[width=0.99\linewidth]{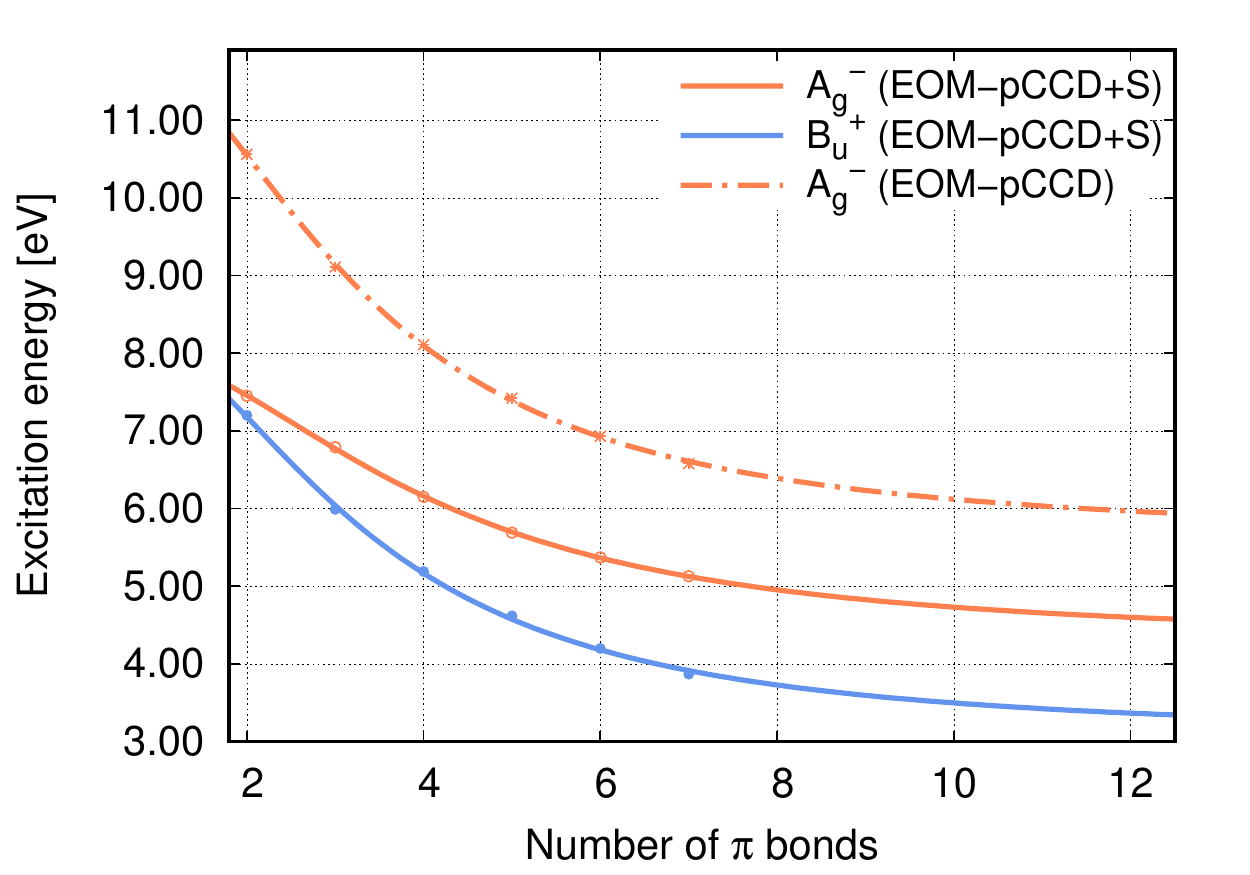}
\caption{Extrapolation of excitation energies calculated for the EOM-pCCD and EOM-pCCD+s models for the first dark state $2^1A_g^-$ (orange line) and the first bright state $1^1B_u^+$ (blue line; EOM-pCCD+S only).
}
\label{fig:excitationsdz}
\end{figure}

We can extrapolate the excitation energies of the two lowest-lying excited states for longer polyenes in the limit of infinite number of $\pi$ bonds and predict the ordering of states for longer polyenes (see, for instance, refs.~\cite{Schulten1976,Tavan1987,Kurashige2004,Starcke2006}).
To do so, the excitation energies can be fitted to a particle-in-a-box model~\cite{Starcke2006} of the form
\begin{equation}
    E(n_\pi) = \frac{a}{L(n_\pi)-b} + c,
\end{equation}
where $L(n_\pi)$ is the length of the polyene that depends on the number of $\pi$ bonds $n_\pi$, while $a,b,c$ are the fitting parameters.
The function for the length of the polyene is obtained by linear regression of the terminal C--C distance of the DFT optimized structures and defines the distance between the terminal carbon atoms as a function of the number of double bonds,
\begin{equation}
   L(n_\pi) = 2.4709 n_\pi - 1.2398.
\end{equation}
The excitation energy at infinite number of $\pi$ bonds is obtained from the fitting parameter $c$.
For EOM-pCCD, the extrapolated excitation energy of the $2^1A_g^-$ state is 5.61 eV and reduces to 4.28 eV for EOM-pCCD+S.
For the $1^1B_u^+$ state, EOM-pCCD+S results in an extrapolated excitation energy of 3.06 eV. 
The resulting fits as well as excitation energies are shown in Figure~\ref{fig:excitationsdz}.
Note that EOM-pCCD+S model overestimates the extrapolated excitation energy of the $1^1B_u^+$ state by approximately 1.30 eV compared to experiment~\cite{polyeneLimit} and by about 0.94 eV compared to MRMP reference data (2.12 eV)~\cite{Kurashige2004}. 

\begin{figure*}[htbp]
\centering
\includegraphics[width=0.99\linewidth]{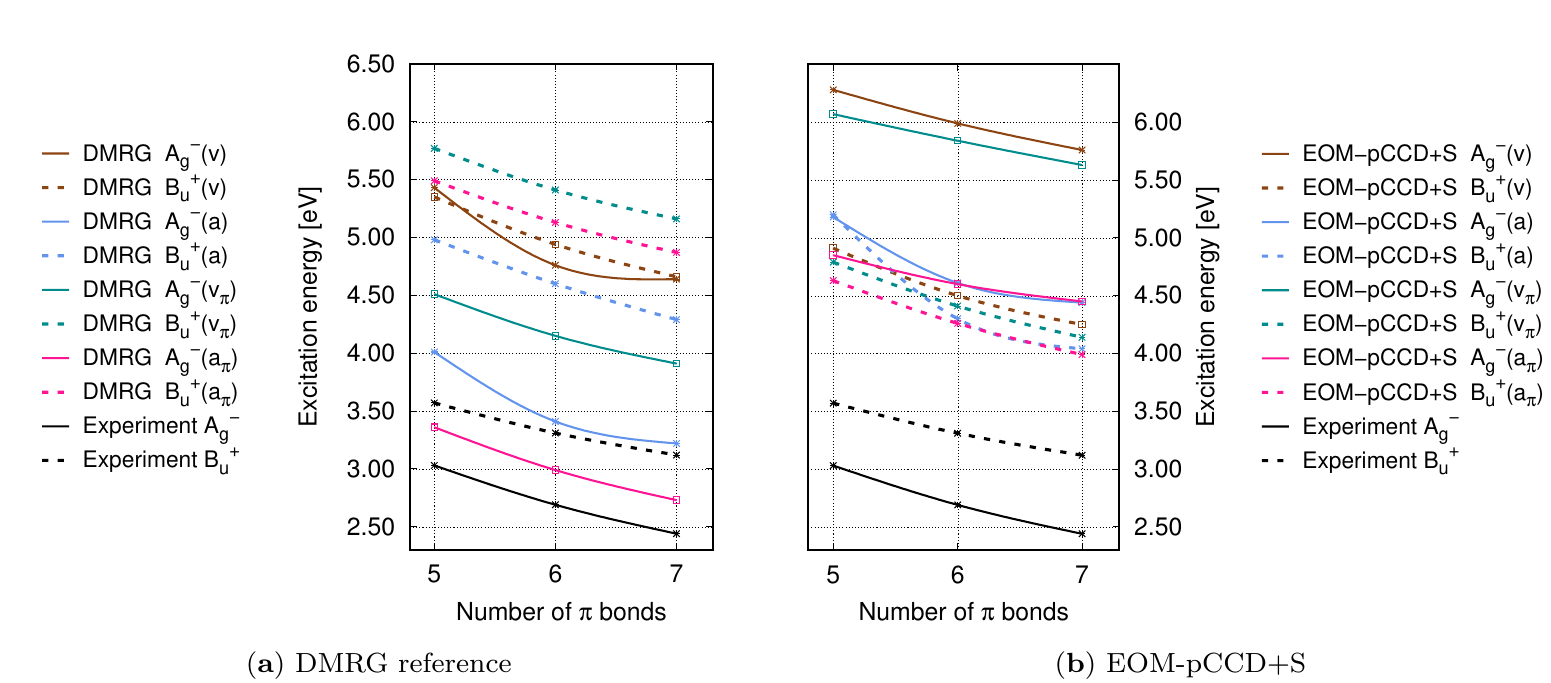}
\caption{Excitation energies in eV of the two lowest-lying excited states for \ce{C10H12}, \ce{C12H14}, and \ce{C14H16} calculated with EOM-pCCD+S.
The molecular structures used in each calculation are given in parentheses and explained in the text (see computational details).
The DMRG reference values are taken from ref.~\cite{DMRG-geom-opt}.
}
\label{fig:dmrg}
\end{figure*}

Finally, we will discuss how the molecular structure affects excitation energies and their character in the EOM-pCCD+S method.
Note, however, that the labels ``$v$'' and ``$a$'' in Table~\ref{tab:dmrg} correspond to the vertically and adiabatically excited state in DMRG calculations.
They do not represent the true adiabatic and vertical excitation energies in EOM-pCCD+S as the DMRG-optimized structures are used.
To simplify the labeling of the excited states and to facilitate direct comparison to DMRG, we will use the same labeling (``$v$'' and ``$a$'') in DMRG and EOM-pCCD+S for all investigated excited states. 
Figure~\ref{fig:dmrg} and Table~\ref{tab:dmrg} show the two lowest-lying excitation energies of \ce{C10H12}, \ce{C12H14}, and \ce{C14H16} for different molecular geometries optimized by DMRG.
For the $2^1A_g^-$ state, the excitation energies predicted by EOM-pCCD+S reduce by approximately 1 eV if the molecular structure is allowed to relaxed (adiabatic excitations).
The choice of the active space in the DMRG structure optimization, however, does not significantly affect the accuracy of EOM-pCCD+S for longer polyene chain lengths (differences are less than 0.1 eV between $v$ ($a$) and $v_\pi$ ($a_\pi$)), while DMRG is more sensitive to the active space (differences in DMRG excitation energies amount to 1 eV).
Similar observation can be made for the $1^1B_u^+$ state.
For relaxed molecular structures, the excitation energies slightly improve by 0.2 eV compared to the vertical excitation energies.
Furthermore, the active space used in DMRG calculations does not change the excitation energies considerably and similar excitation energies are obtained for the vertically and adiabatically excited states, respectively, with EOM-pCCD+S.
We should emphasize that the excitation energies of the $1^1B_u^+$ state predicted by EOM-pCCD+S are closer to experimental values than the corresponding DMRG excitation energies.
Differences between the theoretically determined values and experiment can be partly attributed to the basis set used in calculations as well as missing electron correlation effects not included in the theoretical model.

\begin{figure}[htbp]
\centering
\includegraphics[width=0.99\linewidth]{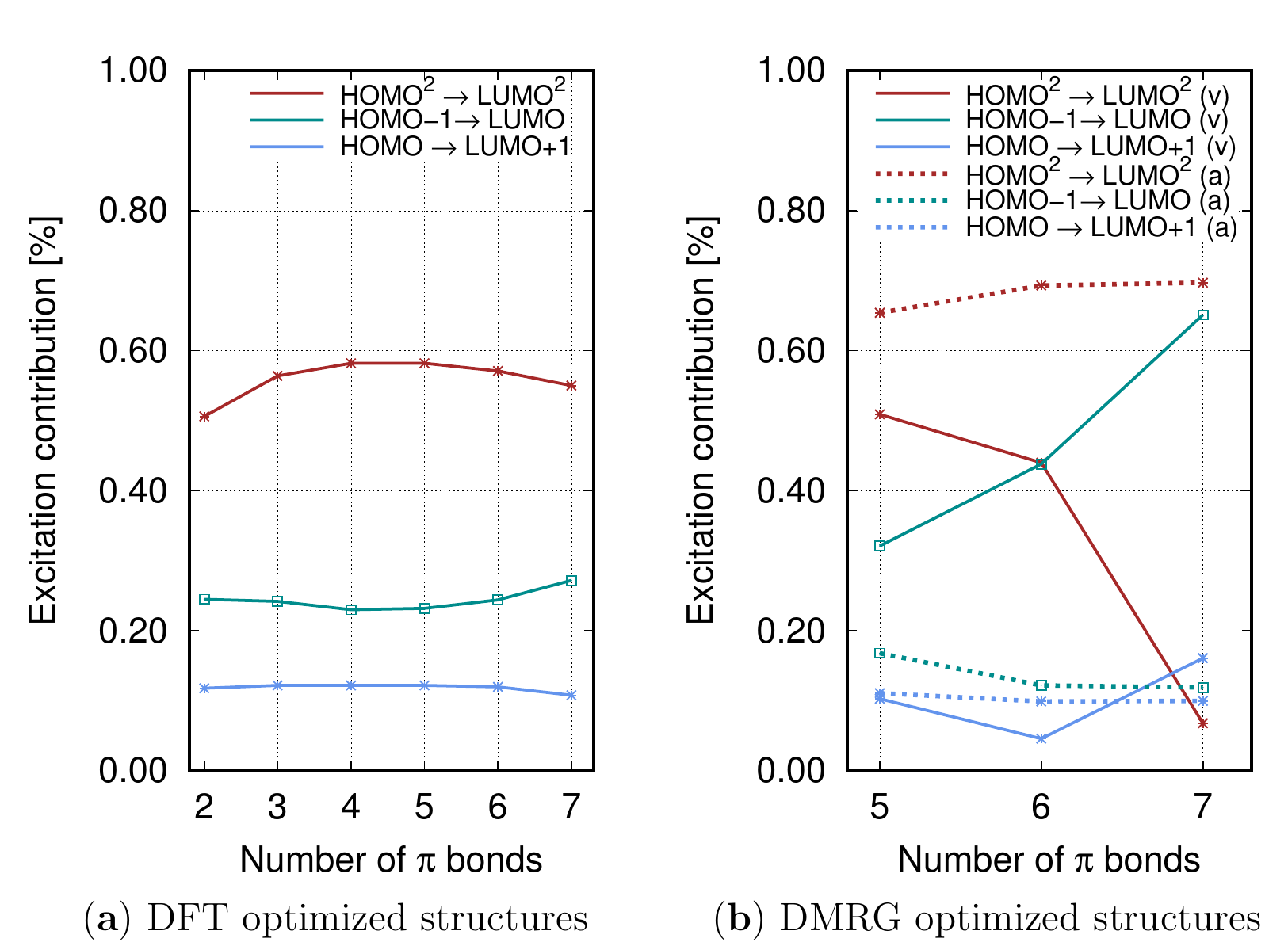}
\caption{Contributions (in percent) of the three dominant configurations of the $2^1A_g^-$ state for the (a) DFT optimized structures (vertical excitations) and (b) DMRG optimized structures (vertical and adiabatic excitations).
}
\label{fig:character}
\end{figure}

Figure~\ref{fig:character} shows the contributions (in percent) of the three dominant configurations of the $2^1A_g^-$ excited state for different molecular geometries.
These are the doubly excited $b_g\rightarrow a_u^*$ (HOMO$^2\rightarrow$LUMO$^2$), the singly excited $b_g \rightarrow b_g^*$ (HOMO$\rightarrow$LUMO+1), and the singly excited $a_u \rightarrow a_u^*$ (HOMO$-$1$\rightarrow$LUMO)~\cite{Schulten1976,Tavan1987}.
For the DFT optimized structures shown in Figure~\ref{fig:character}(a) (vertical excitations), 
the contribution from the doubly-excited configuration HOMO$^2$ $\rightarrow$ LUMO$^2$ increases to approximately 50-60\%, while both singly excited configurations constitute 15 to 25\%. For longer polyene chain lenghts, the contribution from the doubly-excited configuration decreases.

The character of the $2^1A_g^-$ state changes considerably if the molecular structure is modified.
Specifically, the weights of the configurations of the vertically excited $2^1A_g^-$ state strongly depend on the number of $\pi$-bonds if DMRG optimized structures are used (see solid lines in Figure~\ref{fig:character}(b)). While the dark state in \ce{C10H12} is dominated by the HOMO$^2\rightarrow$LUMO$^2$ configuration for the shorter polyene chain lengths, the weight of the HOMO$-$1$\rightarrow$LUMO configuration gradually increases and constitutes more than 60 \% for \ce{C14H16}.
Extending the polyene chain to 7 double bonds further decreases the weight of the HOMO$^2\rightarrow$LUMO$^2$ configuration to less than 10~\%, while the contribution of the singly excited HOMO$\rightarrow$LUMO+1 configuration slightly increases.
Thus the vertically excited $2^1A_g^-$ state of \ce{C14H16} is dominated by only one configuration.

The picture changes completely if the molecular structure of the excited state is allowed to relax (see dashed lines in Figure~\ref{fig:character}(b)).
For growing polyene chains, the weight of the doubly excited HOMO$^2\rightarrow$LUMO$^2$ configuration gradually increases (up to 70~\% for \ce{C14H16}), while both singly excited configurations (HOMO$\rightarrow$LUMO+1) and HOMO$-$1$\rightarrow$LUMO) approach weights of approximately 15~\%.
Although EOM-pCCD+S predicts a dominant contribution of the doubly-excited HOMO$^2\rightarrow$LUMO$^2$ configuration in the $2^1A_g^-$ excited state, it underestimates the overall weight of all doubly-excited configurations ($< 70$ \% in EOM-pCCD+S) that may increase to approximately 80 \% for longer polyene chain lenghts~\cite{Starcke2006}.
This underestimation can be attributed to broken-pair configurations that are not included in the pCCD model and have to be accounted for \textit{a posteriori}~\cite{frozen-pCCD,Kasia-lcc}.
To conclude, EOM-pCCD+S can appropriately describe the three most important configurations present in the dark state (see also ~\cite{Schulten1976,Tavan1987,Starcke2006}) if molecular structures are relaxed.
The qualitatively good description of the character of the first dark state is also evident in the excitation energies.
The adiabatic excitation energies are approximately 1 to 1.5 eV lower in energy than the corresponding vertical ones (compare Tables~\ref{tab:dft} and \ref{tab:dmrg}).

\section{Conclusions}\label{sec:conclusions}

In this work, we have extended the pCCD model to excited states using the equation of motion formalism.
Since the cluster operator of pCCD is restricted to electron-pair states, only electron-pair excitations can be modeled with EOM-pCCD.
To augment the excited-state model with single and general double excitations, we have to modify the cluster operator to include single and general double excitations as well which considerably increases the computational cost compared to the pCCD reference calculation.
To arrive at a cost-effective model similar to pCCD, we have included single excitation \textit{a posteriori} in the CI-type ansatz.
Our method (EOM-pCCD+S) is motivated by the CIS approach that allows us to efficiently model singly excited states with an SCF reference function.
Since only the CI-type ansatz is modified, the computational scaling remains similar to EOM-pCCD (the Hamiltonian contains terms that scale as $\mathcal{O}(o^2v^2)$), but with a larger pre-factor.
Despite the simplicity of the model, EOM-pCCD+S breaks size-intensivity, which can be restored by adjusting the cluster operator.

Although pCCD is commonly combined with an orbital-optimization protocol, we did not optimize the molecular orbital basis in the pCCD reference calculation.
Orbital optimization in pCCD usually results in localized, symmetry-broken orbitals that prevent us from identifying the symmetry of the targeted excited states.
Furthermore, if the orbitals are optimized within pCCD, the optimal set of orbitals is biased toward the ground state and the corresponding excitation energies are generally overestimated.
This bias can be reduced by optimizing the orbitals simultaneously for both ground and excited states, which increases the computational cost of EOM-pCCD and EOM-pCCD+S significantly.
To benefit from the $\mathcal{O}(o^2v^2)$ scaling of both excitation models, we have thus skipped the orbital optimization step and used restricted Hartree--Fock orbitals in all calculations.

Both excitation models have been assessed against excitation energies of the uranyl cation, whose lowest-lying excited states contain purely singly-excited states, and all-trans polyenes containing 2 to 7 $\pi$-bonds.
For the \ce{UO2^{2+}} molecule, EOM-pCCD+S provides excitation energies that are close to CIS results, overestimates, however, the excitation energies of the two lowest-lying excited states by about 0.5 eV compared to EOM-CCSD reference calculations.
For all-trans polyenes, EOM-pCCD+S fails to predict the correct order of the first dark and first bright state, while the excitation energies of the $1^1B_u^+$ state are in good agreement with MRMP results and closer to experiment than DMRG reference data.
Furthermore, the excitation energies and the character of the excited states strongly depend on the molecular structures used in calculations.
Specifically, the character of the $2^1A_g^-$ state can only be properly predicted if molecular structures are allowed to relax, resulting in a dominant doubly-excited HOMO$^2\rightarrow$LUMO$^2$ configuration and two singly-excited configurations (HOMO$\rightarrow$LUMO+1) and HOMO$-$1$\rightarrow$LUMO).
In order to accurately model the first dark state in all-trans polyenes, double excitations beyond electron-pair excitations as well as higher excitations (triples, etc.) might be important to capture the missing correlation effects in the targeted excited states that cannot be described within pCCD.
Possible extension of EOM-pCCD using CC-type corrections on top of the pCCD reference function are presently being developed in our laboratory.
To conclude, EOM-pCCD+S represents a good and cost-effective starting point to investigate singly-excited states in the pCCD model.
Additional numerical studies are currently under investigation in our laboratory.

\section*{Supplementary Material}
See Supplementary Material for excitation energies of selected all-trans polyenes obtained by EOM-pCCD and EOM-pCCD+S in the optimized pCCD orbital basis.

\section*{Acknowledgements}
Financial support from a SONATA BIS grant of the National Science Centre, Poland (no.~2015/18/E/ST4/00584) and a Marie-Sk\l{}odowska-Curie Individual Fellowship project no.~702635--PCCDX is gratefully acknowledged. 

Calculations have been carried out using resources provided by Wroclaw Centre for Networking and Supercomputing (http://wcss.pl), grant No.~412.

We had many helpful discussions with Leszek Meissner, Piotr Piecuch, and Karol Kowalski.
\normalem
\bibliography{rsc} 
\end{document}